\documentclass[aps,prb,twocolumn,groupedaddress,showpacs,superscriptaddress]{revtex4-1}
\usepackage{wasysym}
\usepackage{graphicx}
\usepackage{color}
\usepackage{tabularx}

\begin{document}
\title{The Bose-Hubbard model on a triangular lattice with diamond ring-exchange}

\author{V.G.~Rousseau}
\affiliation{Physics Department, Loyola University New Orleans, 6363 Saint Charles Ave., LA 70118, USA}
\author{K.~Hettiarachchilage, K.-M.~Tam, M.~Jarrell, J.~Moreno} 
\affiliation{Department of Physics \& Astronomy, Louisiana State University, Baton Rouge, Louisiana 70803, USA}
\affiliation{Center for Computation \& Technology, Louisiana State University, Baton Rouge, Louisiana 70803, USA}

\begin{abstract}
Ring-exchange interactions have been proposed as a possible mechanism for a Bose-liquid phase at
zero temperature, a phase that is compressible with no superfluidity. Using the Stochastic Green Function algorithm~(SGF), we study the effect of these interactions for bosons on a two-dimensional triangular
lattice. We show that the supersolid phase, that is known to exist in the ground state for a wide range of densities, is rapidly
destroyed as the ring-exchange interactions are turned on. We establish the ground-state phase diagram of the system, which is
characterized by the absence of the expected Bose-liquid phase.
\end{abstract}

\maketitle

\section*{Introduction}
A prominent direction in atomic physics is to use optically trapped quantum gases to gain insight into many-body interacting problems. Over the last decade or so, the physics of superfluids, Bose-Einstein condensates, and Mott insulators have been demonstrated with $^{87}Rb$ in optical traps. The physics of frustrated systems is among the most active areas in the field of many-body interacting systems. Ring-exchange models, originally proposed for studies of magnetism in Helium-3, have surged recently because of the interest in the study of frustrated spin systems and their relation with gauge theory. \cite{Neto-Pujol-Fradkin-2006,Balents-Fisher-Girvin-2002} In particular, the competition between ring-exchange and conventional hopping terms has been studied intensively.\cite{RingExchange1,RingExchange2,RingExchange3,RingExchange4,RingExchange5,RingExchange6} These studies were motivated by the various exotic phases that were expected to be observed under the influence of  ring-exchange terms. Most notably the hunt for spin liquids, and Bose-liquids--a compressible critical phase without boson condensation-- in a numerically tractable model has been a reigning theme in the simulation of boson and spin models.

Besides the theoretical interest in possible exotic phases, the potential of an experimental realization of boson systems 
with high order ring exchange coupling is tantalizing. Recently, triangular lattices have been realized 
 experimentally by using three intersecting laser beams in a plane \cite{Triangular_lattice}, and kagome lattice has also been proposed \cite{Kagome_lattice}. 
Thus, this opens the avenue for the study of ultracold atoms in frustrated lattices. In addition, the realization of ring-exchange coupling has also
 been proposed theoretically based on the resonant coupling between bosons and a two-particle molecular state for the square lattice \cite{Ring_exchange_lattice}, 
and the strong coupling limit of the Bose Hubbard model for the kagome lattice\cite {Ring_exchange_lattice2}. The first mechanism proposed for the square lattice 
should be generalizable to the triangular lattice provided that the molecular state can be set to reside on each bond of the triangular lattice.

In order to elucidate the effect of ring-exchange processes in a triangular lattice, we perform Quantum Monte Carlo 
simulations for the Bose-Hubbard model with a four-site ring-exchange term with a diamond configuration, as opposed 
to the bow tie configuration in a prior study\cite{Dang-Inglis-Melko-2011}. The model contains a rich phase diagram 
by tuning the four parameters which adjust the hopping, the nearest neighbor repulsion, the ring exchange, and the 
chemical potential. In the classical limit, the model has a massively degenerate ground state. Each ground state 
configuration has one unsatisfied bond at each triangular plaquette. The quantum fluctuation introduced by the hopping 
destroy this degeneracy, and depending upon the filling, either a supersolid or a superfluid phase is formed. On the 
other hand, one can introduce quantum fluctuations via the ring exchange. It has been suggested that the model in this 
limit could become a spin liquid such as that of the $\nu=1/2$ fractional quantum Hall liquid.\cite{Burkov-2010} We 
will demonstrate that the spin liquid phase is preempted by the non-zero superfluid for a very small value of hopping. 
The main effect of the ring exchange term is the suppression of the diagonal ordering in the supersolid. 

The paper is organized as the following. In section II, we discuss the detail of the model and the numerical 
method we used. In contrast to the simulation of the standard Bose Hubbard model, the higher order ring exchange 
leads to a complicated Hamiltonian which is particularly well suited to be solved with the method we 
used--Stochastic Green Function. 
In section III, we discuss the quantities we measure by identifying different phase transitions of the model. We pay 
particular attention to the calculation of the superfluid density on the triangular lattice, which does not have 
orthonormal basis vectors. In the section IV, we provide the numerical results and the phase diagram. Section V is 
the conclusion of the paper.

\section*{Model and numerical method}
\begin{figure}[h]
  \centerline{\includegraphics[width=0.45\textwidth]{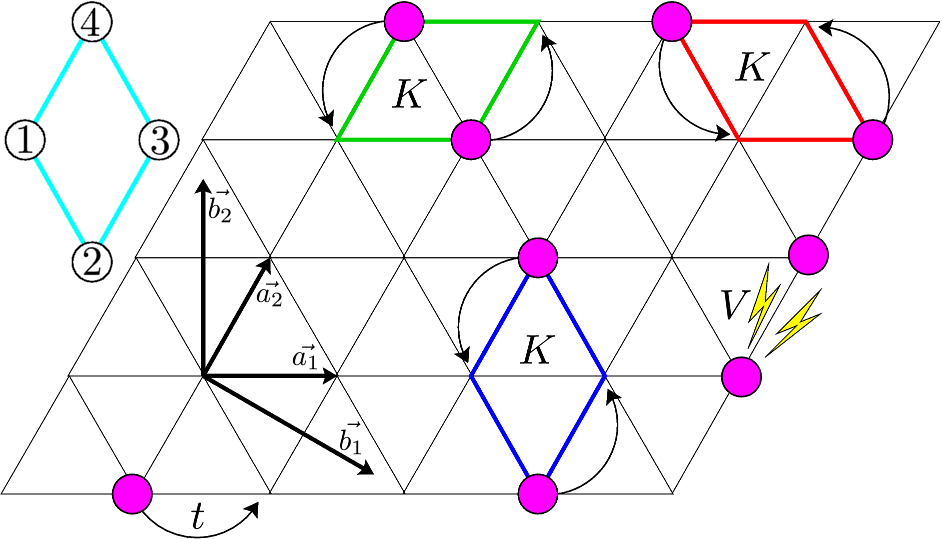}}
  \caption{(Color online) The triangular lattice and the effect of different terms of the Hamiltonian.
  The usual kinetic term $t$ allows the particles to hop between near-neighboring sites. The ring-exchange term $K$ performs a correlated hopping of two particles within the same diamond. This process is possible only if the diamond contains exactly two particles that are either opposite first-neighbors (green diamond) or second-neighbors (blue and red diamonds).  The presence of a pair of first-neighboring particles is penalized by the potential $V$ (yellow). For a given diamond $\diamond$ we label the sites in a counterclockwise fashion, $\diamond 1$, $\diamond 2$, $\diamond 3$, $\diamond 4$, starting from one of the opposite first-neighbors (cyan diamond).}
  \label{TriangularLattice}
\end{figure}

The model we consider consists of hard-core bosons on a two-dimensional triangular lattice (Fig.~\ref{TriangularLattice}). The Bravais lattice is spanned by the basis vectors $(\vec a_1,\vec a_2)$
with lengths chosen as unity, and the reciprocal lattice is spanned by the vectors $(\vec b_1,\vec b_2)$ with lengths $4\pi/\sqrt3$.
The Hamiltonian takes the form (we use periodic boundary conditions)
\begin{eqnarray}
  \nonumber \hat\mathcal H= &-& t\sum_{\langle p,q\rangle}\big(a_p^\dagger a_q^{\phantom\dagger}+H.c.\big)+V\sum_{\langle p,q\rangle}\hat n_p\hat n_q\\
  \label{Hamiltonian}       &-& K\sum_{\diamond}\big(a_{\diamond 1}^\dagger a_{\diamond 3}^\dagger a_{\diamond 2}^{\phantom\dagger} a_{\diamond 4}^{\phantom\dagger}+H.c.\big),
\end{eqnarray}
where $a_p^\dagger$ and $a_p^{\phantom\dagger}$ are the creation and annihilation operators of a hard-core
boson on site $p$, and $\hat n_p=a_p^\dagger a_p^{\phantom\dagger}$ is the number operator on site $p$. The creation and annihilation operators
of hard-core bosons satisfy fermionic anti-commutation rules when acting on the same site,
$a_p^2=0$, $a_p^{\dagger 2}=0$, $\big\lbrace a_p^{\phantom\dagger},a_p^\dagger\big\rbrace=1$, and bosonic
commutation rules when acting on different sites, $\big[a_p^{\phantom\dagger},a_q^{\phantom\dagger}\big]=0$, $\big[a_p^{\dagger},a_q^{\dagger}\big]=0$,
$\big[a_p^{\phantom\dagger},a_q^{\dagger}\big]=0$.
The sums $\sum_{\langle p,q\rangle}$ are over all distinct pairs of neighboring sites $p$ and $q$, and the sum $\sum_\diamond$ is
over all diamonds with all possible orientations (Fig.~\ref{TriangularLattice}, blue, green, red).
The parameter $t$ controls the kinetic energy, and $K$ controls the intensity of the ring-exchange term. Because
ring-exchange processes are possible only if the particles are nearby each other, the ring-exchange term acts as an effective attractive potential and can lead to instabilities\cite{RingExchange2,RingExchange4}. The system is stabilized by the presence of the repulsive potential $V$ between first-neighbors.

In order to solve the model (\ref{Hamiltonian}) we perform quantum Monte Carlo simulations by using the Stochastic Green Function (SGF)
algorithm\cite{SGF} with directed updates\cite{DirectedUpdate}. The SGF method allows us to perform simulations in the canonical
ensemble as well as in the grand-canonical ensemble\cite{GlobalSpaceTimeUpdate}. In the following we take advantage of this flexibility. For simulations in the grand-canonical ensemble we add the usual term $-\mu\hat\mathcal N$ to the Hamiltonian~(\ref{Hamiltonian}) with $\hat\mathcal N=\sum_p\hat n_p$, where the chemical potential $\mu$ allows
us to control the average number of particles.
In order to study the different phases of the system, we measure the dimensionless superfluid density $\rho_s$.
It was recently shown\cite{RousseauSuperfluid} that the well-known formulas that express the superfluid density as a function of
the response of the free energy to a boundary phase twist\cite{Fisher} or to the fluctuations of the winding number\cite{PollockCeperley}
are valid only for a particular class of Hamiltonians. In the case of the Hamiltonian~(\ref{Hamiltonian}), it is easy to see that the
ring-exchange term conserves the center-of-mass of the system, and therefore commutes with the second-quantized position operator. As a result,
the condition~(34) of Ref.\onlinecite{RousseauSuperfluid} is satisfied, allowing the superfluid density to be expressed as a function of the winding number as
\begin{equation}
  \label{NewRhoS} \rho_s=\frac{\big\langle\mathcal W_1^2+\mathcal W_2^2+\mathcal W_1\mathcal W_2\big\rangle}{6t\beta},
\end{equation}
where $\mathcal W_1$ and $\mathcal W_2$ are the winding numbers measured along $\vec a_1$ and $\vec a_2$, and $\beta$ is the inverse temperature (see Eq.~(A12) of Ref.\onlinecite{RousseauSuperfluid}). It is worthwhile to emphasize here that~(\ref{NewRhoS}) is different from the expression\cite{PollockCeperley}
that is sometimes improperly applied to lattices with non-orthonormal primitive vectors. This is due to the fact that the expression
of the Laplacian in non-orthonormal coordinates is associated to a change of the energy scale that must be reflected in the
expression of the superfluid density, and that the non-diagonal metric tensor results in correlations between the winding numbers in
the two primitive directions\cite{RousseauSuperfluid}.
We also measure the static structure factor $S(\vec k)=\big\langle\tilde n(\vec k)^\dagger\tilde n(\vec k)\big\rangle$, with
\begin{equation}
  \tilde n(\vec k)=\frac{1}{L^2}\sum_p\big(\hat n_p-\rho\big) e^{-i\vec k\cdot\vec r_p},
\end{equation}
where $L$ is the linear size of the lattice and $\rho=\big\langle\hat\mathcal N\big\rangle/L^2$ is the dimensionless density of particles. The subtraction of $\rho$ in the above expression is meant to get rid of the Bragg peaks, that is to say $S(\vec K)=0$ for $\vec K=n_1\vec b_1+n_2\vec b_2$
with $n_1,n_2$ integers, and does not affect the value of $S(\vec k\neq \vec K)$.  In order to capture the ground state properties, we use an inverse temperature $\beta=L/t$.

\section*{Analytical preliminaries}
We note that the model~(\ref{Hamiltonian}) is particle-hole symmetric, which allows us to restrict our study to densities $\rho\in[0;\frac12]$. By substituting into (\ref{Hamiltonian}) the creation and annihilation operators of holes,
$h^\dagger_p=a^{\phantom\dagger}_p$ and $h^{\phantom\dagger}_p=a^\dagger_p$, it is straightforward to show that the energy $E(\rho)$ as a function of the density $\rho$ satisfies:
\begin{equation}
  \label{Math1} E(1-\rho)=E(\rho)+3VL^2(1-2\rho)
\end{equation}
By definition, at zero temperature, we have $\mu(\rho)=\frac{\partial E(\rho)}{L^2\partial\rho}$, from which we deduce:
\begin{equation}
  \mu(1-\rho)=6V-\mu(\rho)
\end{equation}
In particular, at half-filling, we have the exact result $\mu(\frac12)=3V$, which is independent of $t$ and $K$.  Since the energy of the system with a single particle is $-6t$, it follows that the system is empty for $\mu<-6t$ and completely
filled for $\mu>6(t+V)$, independent of the value of $K$.

\section*{Numerical results}
Simulations of (\ref{Hamiltonian}) are made difficult by (i) the high number of non-diagonal Hamiltonian terms (six kinetic terms and six ring-exchange terms per site), (ii) the quartic nature of the ring-exchange term that couples four sites and affects two worldlines at a time, and (iii) the geometry of the ring-exchange term that introduces a competition between the formation of pairs of first neighbors and pairs of second neighbors. Nevertheless, we are able to simulate the model with sizes up to $42\times42$.

\begin{figure}[h]
  \centerline{\includegraphics[width=0.5\textwidth]{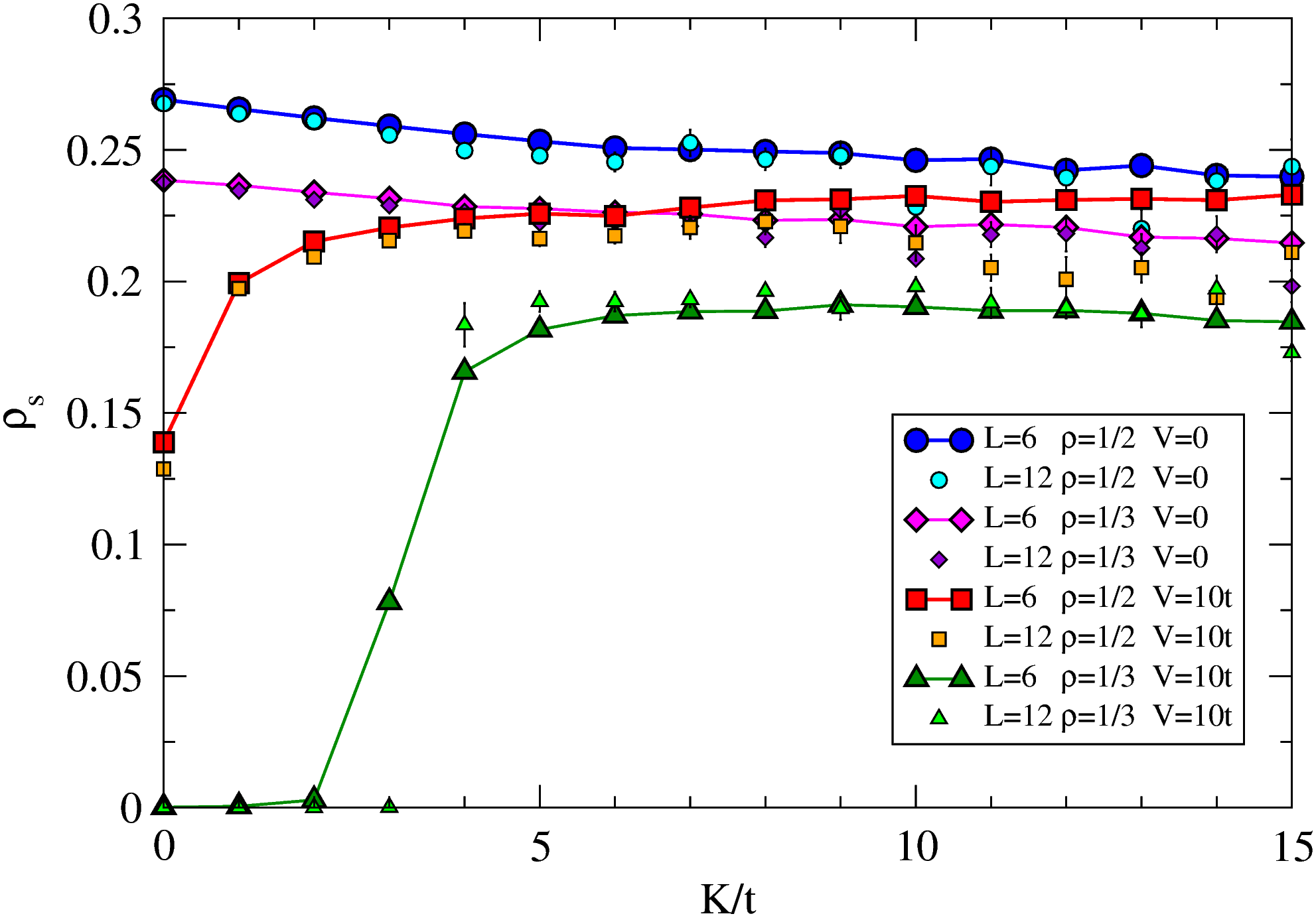}}
  \caption{(Color online) The superfluid density $\rho_s$ as a function of $K/t$ for $V/t=0,10$ and $\rho=\frac13,\frac12$.}
  \label{RhoSvsK}
\end{figure}

Our first main result is that the ring-exchange term does not have the expected effect. It does not destroy superfluidity, at least not for reasonable values of $K$. On the contrary, starting from a solid phase with zero superfluid density, the ring-exchange term breaks the solid order and restores superfluidity as its magnitude is increased. This can be seen in Fig.~\ref{RhoSvsK} which shows the superfluid density $\rho_s$ as a function of $K/t$ in the canonical ensemble for different values of the density $\rho$ and the potential $V/t$. On the one hand, for $V/t=0$, the superfluid density decreases slightly as $K$ increases, as well for $\rho=\frac13$ as for $\rho=\frac12$. On the other hand, for $V/t=10$, the superfluid density increases as the ring-exchange interactions are turned on. We conclude that the system is always superfluid when $K$ is dominant, and that only a competition with $V$ can produce other phases. Therefore, in the remainder of this paper, we work with the fixed potential value $V/t=10$.

\begin{figure}[h]
  \centerline{\includegraphics[width=0.45\textwidth]{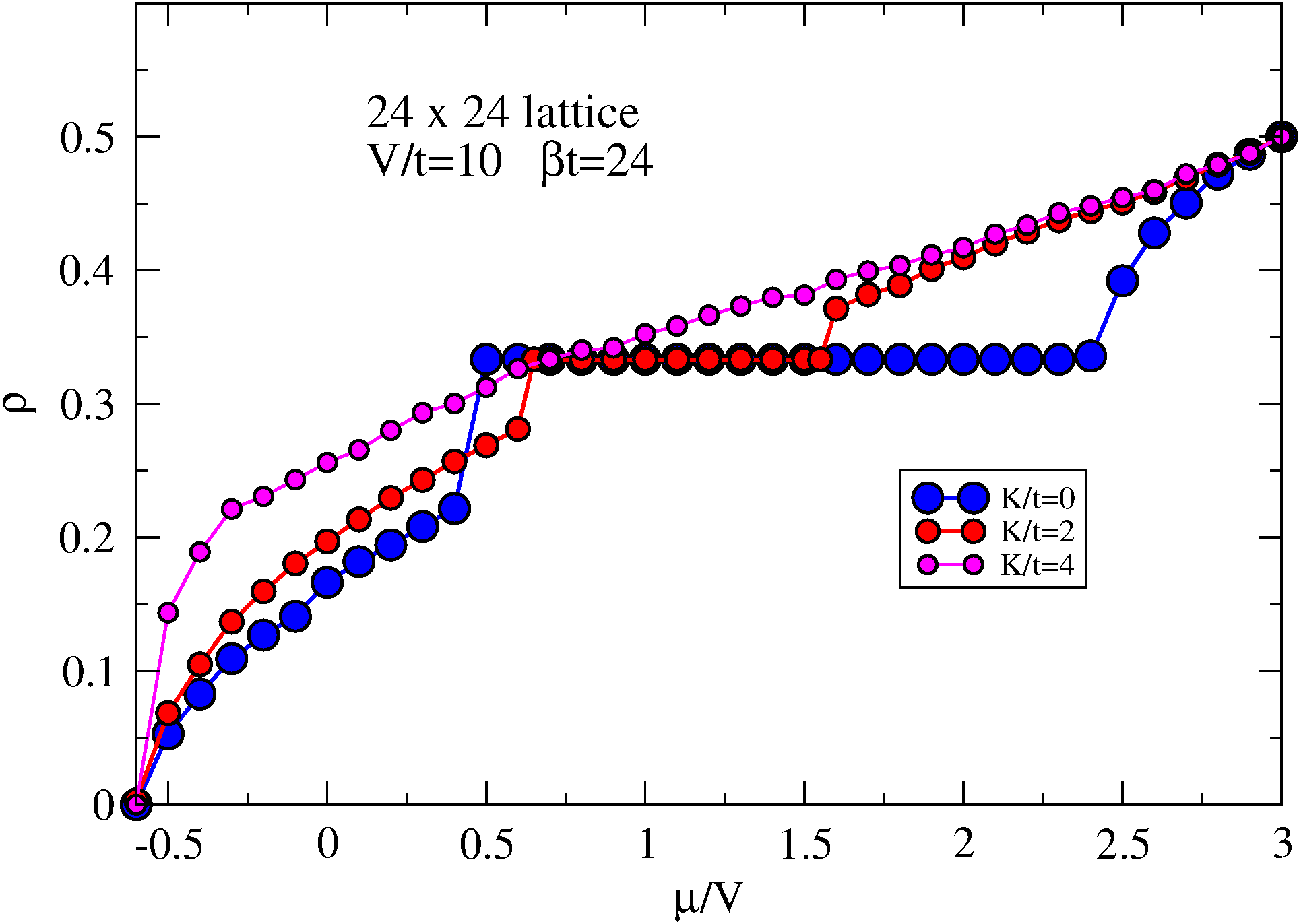}}
  \caption{(Color online) The density as a function of the chemical potential for $V/t=10$ and different values
  of $K/t$. The error bars are about the size of the purple symbols.}
  \label{V10RhoVsMu}
\end{figure}

\begin{figure}[h]
  \centerline{\includegraphics[width=0.5\textwidth]{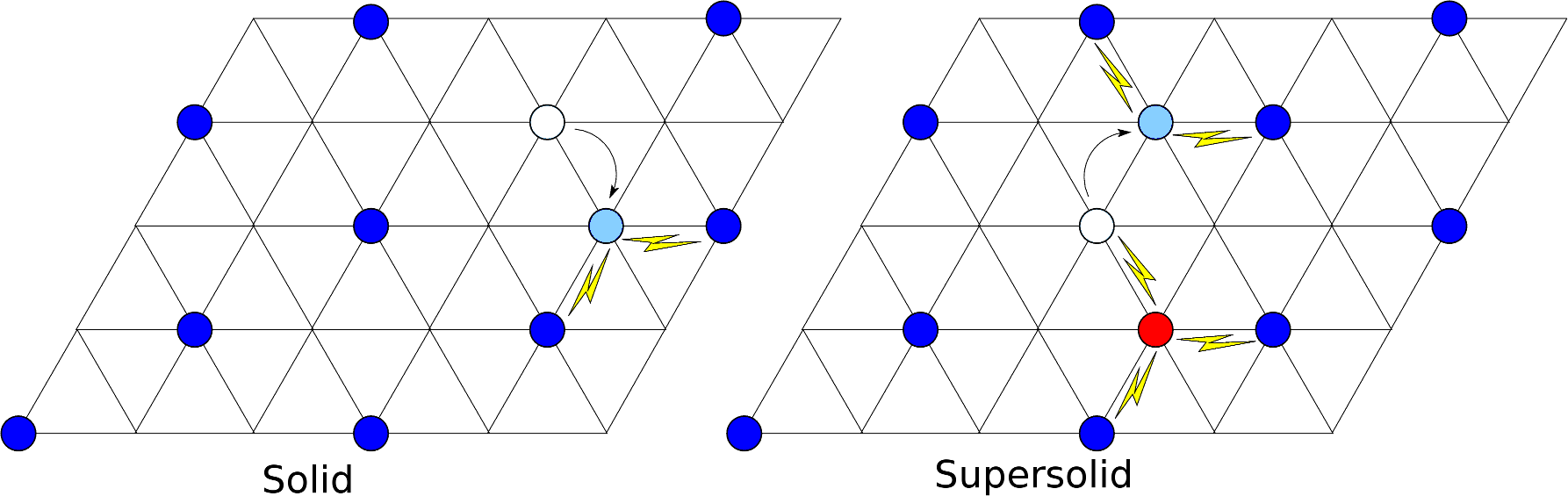}}
  \caption{(Color online) The solid and the supersolid phases. The solid phase ($\rho=\frac13$) is formed by distributing a maximum number of particles without creating pairs of first-neighbors. Any attempt to dislocate one particle (white $\to$ cyan) has an energy cost of $2V$, since two sites would have occupied first neighbors which we term links.  The supersolid phase is obtained by adding extra particles to the solid phase  ($\rho>\frac13$). The presence of an extra particle (red) is associated with the formation of three links with a total energy of $3V$. This extra particle is unable to break the underlying solid structure, because moving one of the extra particle's neighbors (white $\to$ cyan) away would destroy only one link while creating two new extra links, resulting in an energy increase of $V$. However, the extra particle is able to move freely across the solid structure, and can thus
  generate winding associated with superfluidity.}
  \label{Solid-Supersolid}
\end{figure}

It is convenient to start our analysis of the competition between $V$ and $K$ in the grand-canonical ensemble by looking at the density $\rho$ as a function of the chemical potential $\mu$ and different
values of $K/t$~(Fig.~\ref{V10RhoVsMu}).
The slope of these curves, $\frac{\partial\rho}{\partial\mu}$, is proportional to the isothermal compressibility $\kappa_T$. Thus
incompressible (solid) phases are detected by the presence of horizontal plateaus.
The case with no ring-exchange interactions, $K/t=0$, has been extensively studied previously \cite{Murthy,Boninsegni,Melko,Heidarian,Boninsegni2,WesselTroyer}. We reproduce here some of results (blue symbols)
from Wessel and Troyer \cite{WesselTroyer} as a starting point of our study. In agreement with our analytical analysis, particles appear for $\mu/V>-0.6$ and their density increases continuously as the chemical potential is raised.
Around $\mu/V\simeq 0.5$ the density suddenly jumps and remains constant at $\rho=\frac13$, up to $\mu/V\simeq 2.4$. Increasing
the chemical potential further, the density increases continuously and reaches the value $\rho=\frac12$ for $\mu/V=3$.  This is in agreement with the previous study\cite{WesselTroyer} that showed that the system undergoes a first-order phase transition from a superfluid ($\mu/V\lesssim 0.5$) to a solid phase ($0.5\lesssim\mu\lesssim 2.4$), then a second-order phase transition to
a supersolid phase ($\mu/V\gtrsim 2.4$).
The features of the solid phase at $\rho=\frac13$ can be  characterized as a uniform distribution of the particles, thereby avoiding the formation of first-neighboring pairs (Fig.~\ref{Solid-Supersolid}, left). The supersolid phase is a phase in which a fraction of the particles can evolve freely across an underlying solid structure formed by the other particles (Fig.~\ref{Solid-Supersolid}, right).  Further considerations on the structure factor (see below) confirm this.

\begin{figure}[h]
  \centerline{\includegraphics[width=0.5\textwidth]{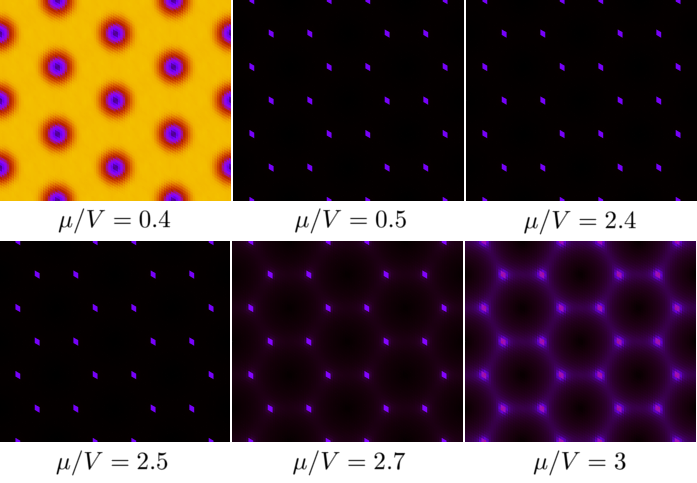}}
  \caption{(Color online) Intensity plots of the structure factor for $K/t=0$ for a $24\times24$ lattice. The nature of the phase transitions can be
  directly seen. The sudden change in the symmetry of $S(\vec k)$ from $\mu/V=0.4$ to $\mu/V=0.5$ is the signature of a
  first-order transition, while the continuous change of the intensity from $\mu/V=2.4$ to $\mu/V=3$ reveals a second-order transition.}
  \label{K00}
\end{figure}

Before discussing the case $K/t\neq 0$, we extend here the previous study\cite{WesselTroyer} by considering intensity plots of the structure factor $S(\vec k)$ (Fig.~\ref{K00}) for $K/t=0$ and different values of the chemical potential, which allow us to directly ``see" the (dis)continuous nature of the transitions.
For $\mu/V=0.4$ the highest intensity (yellow) is small and vanishes in the thermodynamic limit, while the regions with exactly zero-intensity (black) correspond to the locations of the Bragg peaks. Therefore there is no particular structure, as expected
for a superfluid phase. By increasing the chemical potential to $\mu/V=0.5$, the symmetry of the structure factor suddenly changes and peaks, which survive in the thermodynamic limit, appear
at $\vec k=\frac{2}{3}\vec b_1+\frac{1}{3}\vec b_2=\frac{4\pi}{3}\vec a_1$ and symmetry related momenta forming a honeycomb lattice,
in agreement with the expected solid phase (Fig.~\ref{Solid-Supersolid}).
This sudden change is the signature of a first-order transition. The structure factor remains unchanged as the chemical potential
is increased up to $\mu/V=2.4$. Raising the chemical potential further, the density starts to increase again but the symmetry of
the structure factor remains unchanged, with a continuous change of the intensity that results in the peaks becoming smoothly ``linked" to each other. This continuous change of the structure factor from the solid phase to the supersolid phase illustrates the second-order nature of the phase transition.

\begin{figure}[h]
  \centerline{\includegraphics[width=0.5\textwidth]{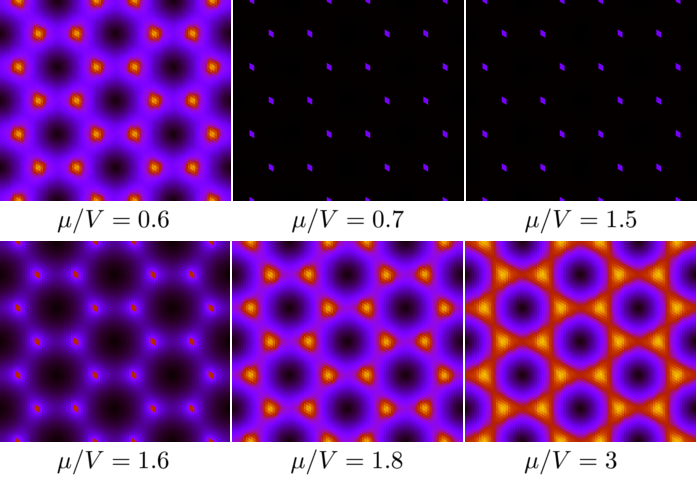}}
  \caption{(Color online) Structure factor for $K/t=2$ for a $24\times24$ lattice. The sudden change in the intensity of $S(\vec k)$
  from $\mu/V=0.6$ to $\mu/V=0.7$ and from $\mu/V=1.5$ to $\mu/V=1.6$ is the signature of a first-order transition.}
  \label{K02}
\end{figure}

We now turn on the ring-exchange interactions and consider again the density $\rho$ as a function of the chemical potential $\mu$
(Fig.~\ref{V10RhoVsMu}) for the case $K/t=2$ (red symbols). We notice that the width of the plateau at $\rho=\frac13$, where
the phase is incompressible, decreases and extends now from $\mu/V\simeq 0.7$ to $\mu/V\simeq 1.5$. Also, raising the chemical
potential further to $\mu/V=1.6$ leads to a sudden jump of the density, suggesting that the previously observed second-order
phase transition is now replaced by a first-order phase transition. Taking a look at the structure factor confirms this scenario
(Fig.~\ref{K02}). Starting from the incompressible phase, $0.7\leq\mu/V\leq1.5$, decreasing the chemical potential
to $\mu/V=0.6$ or increasing it to $\mu/V=1.6$ leads to a sudden change of the intensity and the appearance of ``links" between
the peaks. Thus, the discontinuity associated with the first-order transition is again directly reflected in the structure factor.
The finite compressibility and the peaks in the structure factor for densities $\rho\neq\frac13$ suggest that the phase might be
supersolid. A finite-size scaling below shows that, at half-filling, the supersolid phase survives only at small $K/t$.

\begin{figure}[h]
  \centerline{\includegraphics[width=0.5\textwidth]{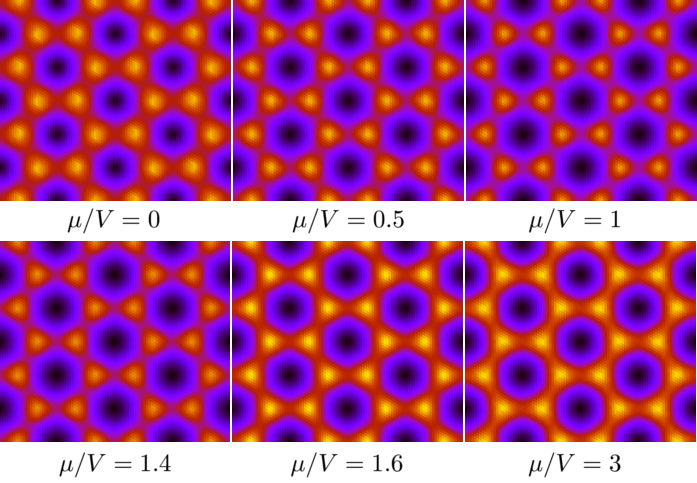}}
  \caption{(Color online) Structure factor for $K/t=4$ for a $24\times24$ lattice. The symmetry of $S(\vec k)$ remains the
  same for all fillings, with a slight dependence in $\mu$ of the intensity.}
  \label{K04}
\end{figure}

Going back to Fig.~\ref{V10RhoVsMu}, we analyze now the case $K/t=4$ (purple symbols) for which the previously observed plateau at $\rho=\frac13$ is absent. The compressibility remains finite for all fillings, with no discontinuity in the density. A consequence is that the width of the plateau observed at $K/t<4$ can be continuously decreased to zero by increasing $K$ up to $K/t=4$. This suggests that the symmetry of the phase should remain the same at all fillings, and identical to the phase present à $K/t=2$ and $\rho>\frac13$. This is confirmed by looking at the structure factor which shows a slow dependence of the intensity as a function of the chemical potential, but no change in the symmetry (Fig.~\ref{K04}).

\begin{figure}[h]
  \centerline{\includegraphics[width=0.5\textwidth]{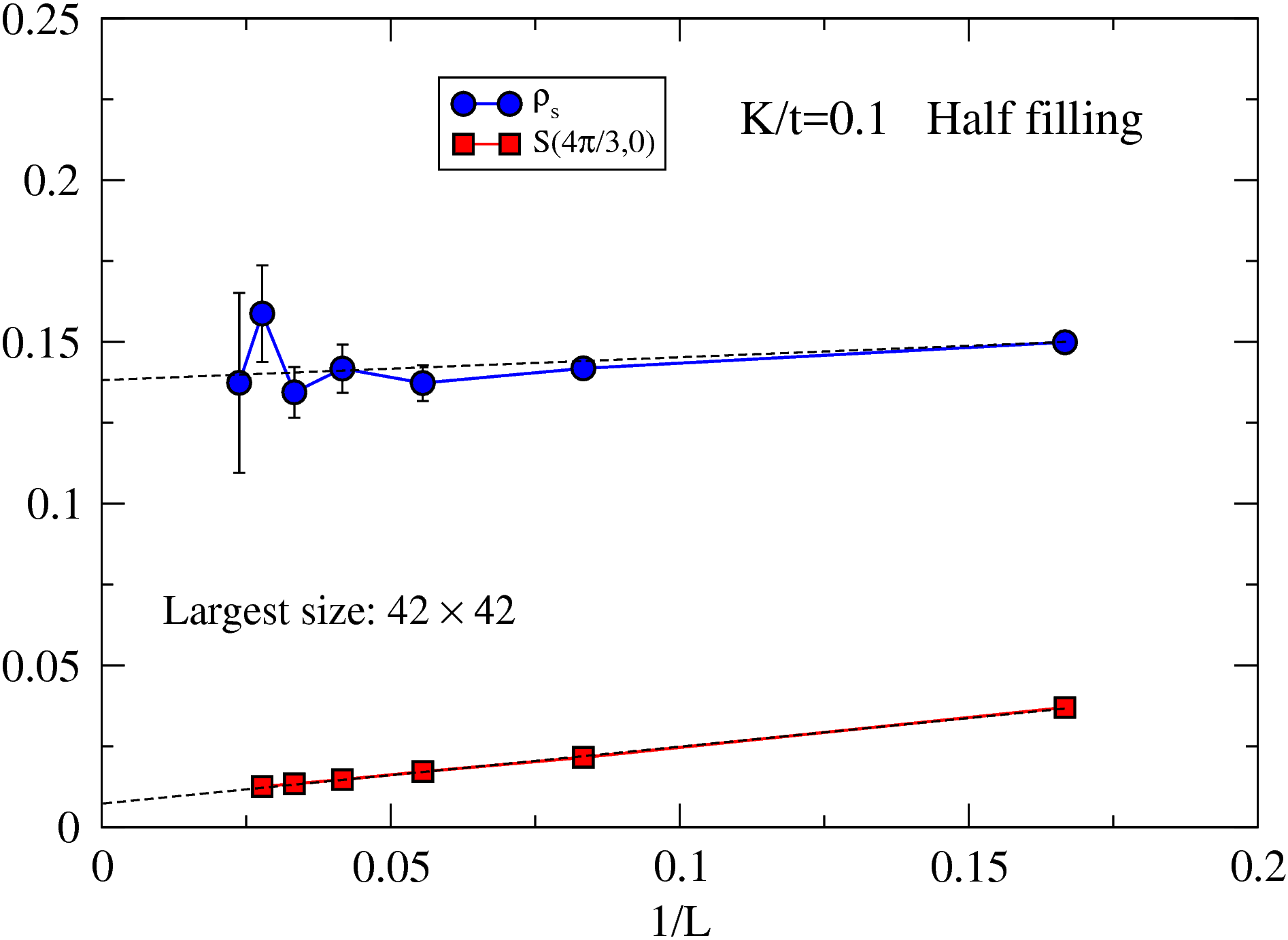}}
  \caption{(Color online) The superfluid density and the structure factor as functions of the inverse system size, for sizes up to
  $42\times42$ and $K/t=0.1$ at half-filling.   Both quantities extrapolate to a finite value in the thermodynamic limit, which is
  the signature of the supersolid phase.}
  \label{Scaling1}
\end{figure}

\begin{figure}[h]
  \centerline{\includegraphics[width=0.5\textwidth]{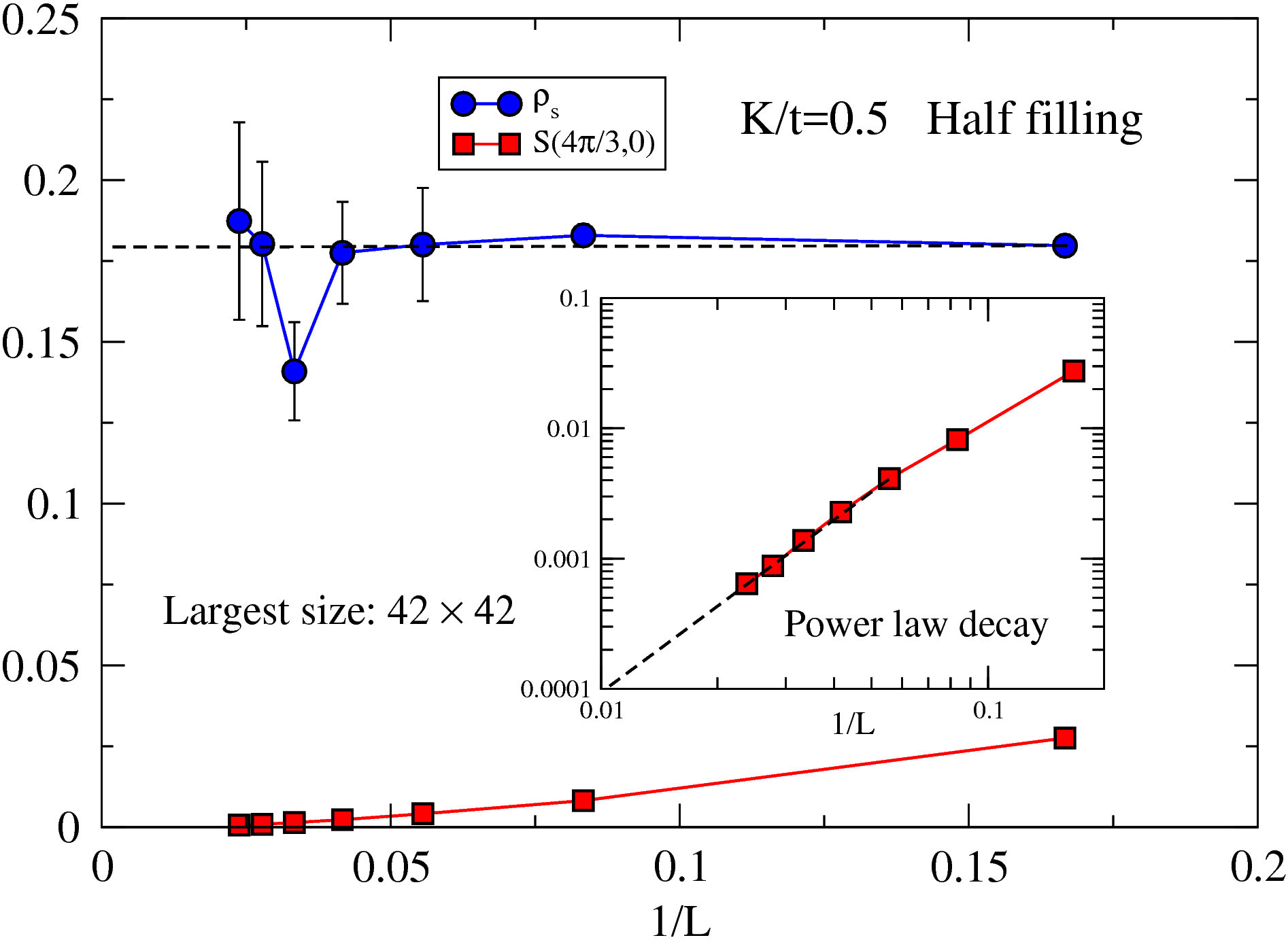}}
  \caption{(Color online) The superfluid density and the structure factor as functions of the inverse system size, for sizes up to
  $42\times42$ and $K/t=0.5$ at half-filling. While the superfluid density extrapolates to a finite value in the thermodynamic
  limit, the structure factor decays as a power law yielding a superfluid phase.}
  \label{Scaling2}
\end{figure}

In order to confirm that the supersolid phase survives for non-zero $K$ values, it is necessary to analyze the scaling of the superfluid density and the structure factor as the size of the system increases. This is conveniently done in the canonical ensemble, where
we can set the density $\rho$ constant and vary the size of the system.  Fig.~\ref{Scaling1} shows $\rho_s$ and $S(4\pi/3,0)$ as functions of $1/L$ for sizes up to $42\times42$ and $K/t=0.1$ at half-filling. The data clearly show that both the superfluid
density and the structure factor converge to finite values, which is the signature of the supersolid phase (Fig.~\ref{Solid-Supersolid}).
However, for $K/t=0.5$ (Fig.~\ref{Scaling2}), a log-log plot of the structure factor (inset) reveals a power law decay in the thermodynamic limit, where only the superfluid density remains finite. As a result, at half-filling, there exists a critical value
$K_c$ for the ring-exchange parameter above which the underlying solid structure of the supersolid phase is destroyed, leading
to a phase transition to a superfluid. By performing plots similar to Fig.~\ref{Scaling1} and Fig.~\ref{Scaling2} (not shown), we find
that $K_c/t=0.35\pm0.05$.

\begin{figure}[h]
  \centerline{\includegraphics[width=0.45\textwidth]{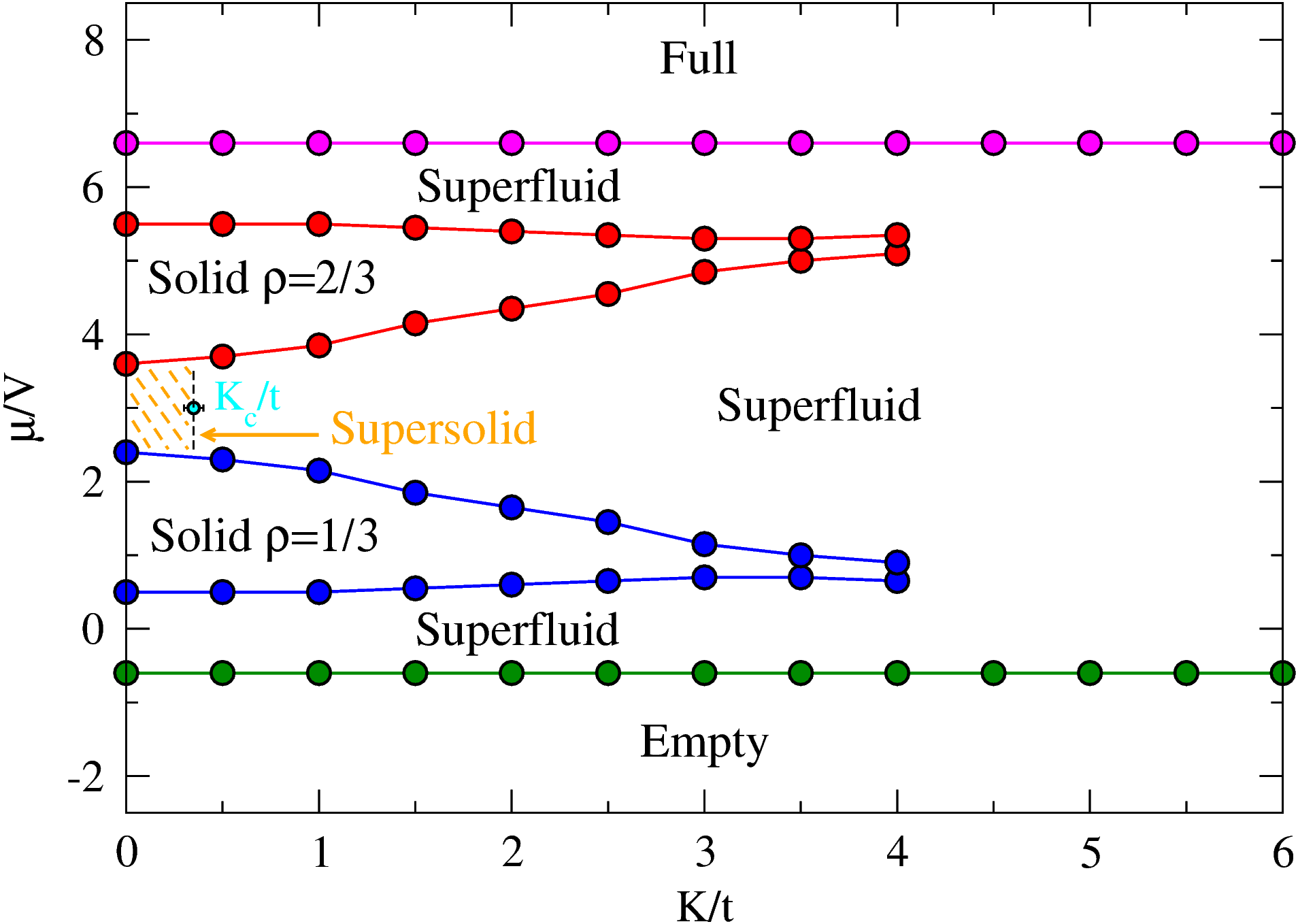}}
  \caption{(Color online) The zero-temperature phase diagram for $V/t=10$. The border (dashed line) between the supersolid and superfluid regions
is symbolic, only the point $K_c$ on this border is obtained from QMC simulations. The exact shape will be determined in further work.}
  \label{PhaseDiagram}
\end{figure}

Finally, by performing simulations in the canonical ensemble, we are able to easily determine the boundaries of the solid phase and draw the zero-temperature phase diagram, Fig.~\ref{PhaseDiagram}. Note that the border (dashed line) between the supersolid and superfluid regions is symbolic, only the point $K_c$ on this border is obtained from QMC simulations. The exact shape will be determined in further work.

\section*{Conclusion}
We study the hard-core Bose Hubbard model on a triangular lattice with four-site ring-exchange terms using the Stochastic Green Function algorithm.  We revise the results from the simulation of the model without the ring exchange using a corrected formula for the superfluid density on a triangular lattice. We then explore a large range of parameters to map out the ground state phase diagram, and find that the system contains three phases, superfluid, solid, and supersolid. 

In the limit of zero hopping, the model has been proposed as a candidate of a spin liquid phase with the character of interacting bosons in the lowest Landau level. While both the ring exchange and hopping terms provide different quantum fluctuations, without hopping not all configurations are possible so the system is not ergodic at zero temperature. So quantum Monte Carlo will not lead into a truly equilibrium phase by the ring exchange and the diagonal interaction terms alone. The result depends on the initial condition of the system. Specifically, the superfluid density or equivalently the winding number can be shown to be zero, thus ruling out the superfluid phase with long range off-diagonal ordering. We believe this is not a physically admissible state at zero temperature due of the lack of ergodicity.

We thus study the model with a finite hopping. We find that the supersolid phase, which is known to exist in the ground state of the $K=0$ model for a wide range of densities, is rapidly destroyed as the ring-exchange interactions are turned on. The solid backbone of the supersolid phase contains two particles and one particle at each triangular plaquette for $2/3$ and $1/3$ filling, respectively, due to the diagonal density-density repulsion. These configurations are not compatible with the off-diagonal ring exchange term.  The ring exchange term favors the oscillations between the two particles and one particle configuration in a triangular plaquette.  The off-diagonal ring exchange interaction disrupts the configuration favored by the diagonal density-density interaction and thus suppresses the solid ordering. 

We establish the ground-state phase diagram of the system, which is characterized by the absence of the Bose-liquid phase when the hopping is non-zero. All the phases possess either diagonal solid ordering, off-diagonal ordering with boson condensation, or both of them. 

\begin{acknowledgments}
This material is based upon work supported by the National Science Foundation under the NSF EPSCoR Cooperative 
Agreement No. EPS-1003897 with additional support from the Louisiana Board of Regents (MJ and KMT), and by 
NSF OISE-0952300 (VGR, KH and JM).
\end{acknowledgments}

\end{document}